\documentclass[12pt,preprint]{aastex}
\newcommand{\Msun}{\ifmmode\mbox{M}_{\odot}\else$\mbox{M}_{\odot}$\fi}
\newcommand{\Rsun}{\ifmmode\mbox{R}_{\odot}\else$\mbox{R}_{\odot}$\fi}
\newcommand{\Mearth}{\ifmmode\mbox{M}_{\oplus}\else$\mbox{M}_{\oplus}$\fi}
\newcommand{\Rearth}{\ifmmode\mbox{R}_{\oplus}\else$\mbox{R}_{\oplus}$\fi}
\newcommand{\chandra}{\textit{Chandra}\,}

\newcommand{\xmm}{\textit{XMM}\,}
\newcommand{\xte}{\textit{RXTE}\,}

\newcommand{\seven}{RXS~J170849.0$-$400910\,}

\newcommand{\etff}{AX~1845$-$0258}
\newcommand{\ttfn}{1E~2259$+$586}
\newcommand{\tfe}{1E~1048.1$-$5937\,}
\newcommand{\etten}{XTE~J1810$-$197\,}
\newcommand{\ninet}{SGR~1900$+$14}

\shorttitle{1E 2259+586 radiative evolution}
\shortauthors{Zhu, Kaspi, Dib,  Woods , Gavriil \& Archibald}

\begin{document}
\title{The Long-term Radiative Evolution of Anomalous X-ray Pulsar 1E~2259$+$586 after its 2002 Outburst}
\author{Weiwei Zhu\altaffilmark{1},
Victoria M. Kaspi\altaffilmark{1,2},
Rim Dib\altaffilmark{1},
Peter M. Woods \altaffilmark{3,4},
Fotis P. Gavriil \altaffilmark{5,6} and
Anne M. Archibald \altaffilmark{1}} 
\altaffiltext{1}{\footnotesize Department of Physics,
McGill University, Montreal, QC, H3A 2T8, Canada;
zhuww@physics.mcgill.ca, vkaspi@physics.mcgill.ca, rim@physics.mcgill.ca,
aarchiba@physics.mcgill.ca}
\altaffiltext{2}{Canada Research Chair; Lorne Trottier Chair;
  R. Howard Webster Fellow of CIFAR}
\altaffiltext{3}{Dynetics, Inc., 1000 Explorer Boulevard, Huntsville,
  AL, 35806}
\altaffiltext{4}{NSSTC, 320 Sparkman Drive, Huntsville, AL, 35805}
\altaffiltext{5}{NASA Goddard Space Flight Center, Astrophysics
  Science Division, Code 662, Greenbelt, MD, 20771} 
\altaffiltext{6}{Center for Research and Exploration in Space Science and
Technology, University of Maryland Baltimore County, 1000 Hilltop Circle,
Baltimore, MD 21250}

\begin{abstract}
We present an analysis of five X-ray Multi-Mirror Mission (\xmm)
observations of the anomalous X-ray pulsar (AXP) 1E 2259+586 taken in 2004
and 2005 during its relaxation following its 2002 outburst.
We compare these data with those of five previous \xmm observations taken in 2002 and 2003, and find the observed flux decay is well described by a power law of index $-0.69\pm0.03$.
As of mid-2005, the source may still have been brighter than preoutburst, and was certainly hotter.
We find a strong correlation between hardness and flux, as seen in other AXPs.
We discuss the implications of these results for the magnetar model.

\end{abstract}


\section{Introduction}
\label{sec:intr}
It is now commonly believed that soft gamma-ray repeaters (SGRs) and anomalous
X-ray pulsars (AXPs) are neutron stars with ultra-strong magnetic fields, i.e.
magnetars \citep{dt92a}.  Their common nature was conclusively demonstrated
when AXP \tfe was observed to emit SGR-like bursts in 2001 \citep{gkw02} and \ttfn, in the supernova remnant (SNR) CTB 109, was seen to undergo a major SGR-like outburst in 2002 \citep{kgw+03,wkt+04}.
Subsequently, a variety of different types of activity in AXPs have been seen,
including short- and long-term flux variations \citep{gk04,dkg07} and slow and rapid pulse profile changes \citep{ikh92,kgw+03,wkt+04,icd+07,dkg07,dkg07b}, in addition to bursts and outbursts (\citealp{gkw06,wkg+05,dkg07}; see \citealt{kas07} for a recent review).

During \ttfn's 2002 outburst, the pulsed and persistent fluxes rose suddenly by a factor of $\ge$20 and decayed on a timescale of months.
Coincident with the X-ray brightening, the pulsar suffered a large glitch of fractional frequency change $4\times10^{-6}$ \citep{kgw+03,wkt+04}. 
In the first few hours of the outburst, the pulsar's pulse profile changed significantly, its pulsed fraction decreased, and its spectrum hardened dramatically. 
Over 80 short SGR-like bursts from the pulsar were observed at the same time \citep{gkw04}.
A near-infrared (K$_s$) enhancement was also observed during the epoch of the outburst \citep{kgw+03}.

Combining {\it Rossi X-Ray Timing Explorer } (\xte) observations and \xmm
observations of \ttfn\, taken during and after the outburst, \cite{wkt+04}
found that the decay of \ttfn's unabsorbed flux (mostly inferred from \xte
pulsed fluxes) after the outburst was well characterized by two power law
components: a rapid steep decay visible only during the first several hours
($<1$ day) of the outburst, and a slower decay of index $-0.22$ for the next several months.
\cite{tkvd04} found that the near-infrared enhancement at late times decayed at the same rate as the slow X-ray decay,
although there were no IR observations during the first few hours of the outburst.

Other AXPs have also exhibited transient behavior that could be explained by SGR-like outbursts. AXP \etten is called transient because it was only discovered in 2003 when it suddenly became brighter by a factor of 100 \citep{ims+04,ghbb04}. 
\cite{gh07} found that the flux of \etten after 2003 followed an exponential decay of timescale 233.5 days. 
Similarly, the AXP CXOU~J164710.2$-$455216 was found to have brightened by a factor of $\sim$300 between two \xmm observations taken 5 days apart in 2006 September \citep{icd+07}.
Candidate AXP \etff, was discovered in an observation made in 1993 by {\it ASCA} \citep{gv98,tkk+98}. Follow-up observations in 1999 showed that the source's flux was smaller by a factor of $\sim$10 \citep{vgtg00}. 
\cite{tkgg06} found that \etff\, remains undetected in \chandra observations
taken in 2003, with its flux $\sim $260-430 times fainter than observed in 1993.  

The transient AXP phenomena summarized above are qualitatively similar to the 1998 August 27 flare of \ninet, in which the X-ray flux decayed with a power law of index $\sim -0.9$ \citep{fmw+03}, and the flux decay of SGR 1627$-$41 since 1998, which followed a power law of index $\sim -0.47 $ and lasted for $\sim$800 days \citep{kew+03}.
However, thus far,  the AXP outbursts have been much less energetic than most SGR outbursts. 
Also, most of the burst energy was released during the afterglows of the AXP outbursts, while for SGR outbursts, the X-ray afterglows have less integrated energy than the burst itself.

With now a handful of AXP and SGR outbursts and subsequent relaxations observed, we can begin to look for correlations between different outburst and relaxation properties in the hope of constraining magnetar physics.
For example, SGR outburst recoveries have been modeled as crustal cooling following impulsive heat injection, and in principle can yield constraints on the nature of the crustal matter \citep{let02}.
Alternatively, the AXP events have been interpreted in terms of magnetospheric twisting \citep{tlk02,bt07}, whose recovery depends on electrodynamics in the region of the
magnetosphere immediately outside the stellar surface.
On the other hand, \citet{gogk07} suggest that AXP recoveries can be
modeled with a stationary magnetosphere, with only the surface temperature
changing.  They argue that their model, which includes the stellar
atmosphere, can be used to quantitatively determine the source's magnetic
field.

In this paper we present a spectral and pulsed flux analysis of 10 \xmm\, observations of AXP 1E~2259+586\ taken between 2002 and 2005, as the source relaxed back toward quiescence following its 2002 outburst.
We compare the X-ray flux and spectral evolution of \ttfn\, with those of other magnetars, and interpret these results in terms of the magnetar model.

\section{Observations}
\subsection{{\textit XMM-Newton} Observations}
\label{sec:xmmobs}
Ten \xmm \citep{jla+01} observations were analyzed for this paper.
The first five observations of \ttfn\, were taken between 2002 and 2003, just prior to and after the 2002 June outburst.
Data from these five observations have already been presented in \cite{wkt+04}.
We re-analyzed these observations using the \xmm calibrations published on
2007 September 4 (XMM-CCF-REL-239\footnote{See http://xmm.esac.esa.int/external/xmm\_sw\_cal/calib/rel\_notes/index.shtml}).
The later five observations were taken between 2004 and 2005. 
Most of these observations pointed at \ttfn,  with the European Photon Imaging Camera (EPIC)  pn camera \citep{sbd+01} in Small Window Mode.
However, three observations were obtained with \xmm pointing at a portion of the SNR CTB 109's shell and with the pn camera in extended Full Frame Mode.
Details about the observational modes, pointing offsets, and exposure times are presented in Table \ref{tab:obs}.
The EPIC mos cameras \citep{taa+01} were operating in Full window mode with
the medium filter in four of the first five observations, the exception being
the third, and therefore the observed spectra are highly piled-up. The mos
cameras were operated in Small Window Mode with the thick filter in the
remaining observations, and hence with lower efficiency than for the pn camera.
Nevertheless, we analyzed the mos data and found the resulting fluxes and
parameters were quantitatively in agreement with those from pn data, given the
current knowledge of cross-calibration uncertainties between the two
instruments.\footnote{See
http://xmm.esac.esa.int/docs/documents/CAL-TN-0018-2-6.pdf, on the current
calibration status of the EPIC cameras.}
In this paper we report only the higher quality EPIC pn data.

The data were analyzed with the \xmm Science Analysis System (SAS) version
7.1.0\footnote{See http://xmm.esac.esa.int/sas/7.1.0/} and the latest calibrations.
Strong background flares can sometimes contaminate source events.
To exclude possible flares, we extracted light curves from the entire field of
view for events having energy $>$ 10 keV.  
We then examined these light curves for flares.  We defined bad time intervals to be when flares occurred, and excluded these intervals for all subsequent analyses.  
For all the \xmm observations, we filtered a total of 20 ks of bad time intervals.
Then we corrected the event times to the barycenter using the SAS {\tt barycen } tool.

\subsection{\textit{RXTE} observations}
\label{sec:xteobs}
We have observed AXP 1E~2259+586 regularly since 1997 with \xte 
(see, e.g., \citealt{gk02}).  Our data were obtained using the Proportional
Counter Array (PCA) on board \xte \citep{jmr+06}. The PCA consists of five identical
and independent xenon/methane Proportional Counter Units (PCUs). We use our
\xte observations of 1E~2259+586 to monitor its pulsed flux, 
and its frequency evolution using phase-coherent timing, and to look for
bursts and pulse profile changes.

For the purposes of this paper we analysed 193 observations that took place
between 2001 April~1 (MJD 52,000) and 2006 September~22 (MJD 54,000): 15
preoutburst observations, 1 observation during the outburst, and 177
postoutburst observations. All 193 observations with the exception of the
two observations immediately following the outburst were taken in {\tt
GoodXenonwithPropane} or {\tt GoodXenon} data modes. Both data modes record
photon arrival times with 1 $\mu$s resolution and bin photon energies into
one of 256 channels. To maximize the signal-to-noise ratio, we analysed only
those events from the top xenon layer of each PCU. The remaining two
observations were in event modes with a time resolution of
$\sim$125~$\mu$s, a smaller number of energy channels, and no possibility of
layer selection.
For each of the observations we created barycentered light curves in the
2--10~keV band with 31.25~ms time resolution. 

We then folded each of the
light curves using an ephemeris determined iteratively by maintaining phase
coherence (see, e.g., \citealt{gk02}). We then used the folded profiles to 
calculate the pulsed flux for each observation using both an rms estimator
(see, e.g., \citealt{wkt+04})  and an area estimator after baseline subtraction (see
\citealt{adk08} in preparation, for details). The results obtained using the two
methods were consistent with each other. Here we only report the area
pulsed flux because, while more sensitive to noise, 
it is the quantity of primary interest.

To calculate the area pulsed flux for a given folded time series, we used
the following:
\begin{equation}
\label{eq:pf}
PF_{area} = \sum\limits_{i=1}^{N} (p_i - p_{min})/N,
\end{equation}
where $p_i$ refers to the count rate in the $i$th bin, $N$ is the number of 
phase bins, and $p_{min}$ is the average count rate in the off-pulse phase of the profile, determined by
cross-correlating with a high signal-to-noise ratio template, and calculated in
the Fourier domain after truncating the Fourier series to six harmonics.
Finally, we combined the pulsed flux numbers from each of two consecutive weeks
into a single number, with the exception of the burst observation and the
two observations that followed it, which remained unbinned. The results are
presented in Figure \ref{fig:all}{\em a}.

\section{Analysis and Results}
\label{sec:res}

\subsection{Spectrum evolution}
Source spectra were extracted from circular regions of 32$''$.5 radius around
the source center, using the barycentered, filtered event file described in
\S \ref{sec:xmmobs}. 
Background spectra were extracted from circular regions of 50$''$  radius centered  $\sim 3'$ away from the source center.
For the observations taken in Small Window Mode, we extracted single- and double-photon events and excluded events on or close to a bad pixel using the filter expression ``FLAG $=0$ \&\& PATTERN $<=4$''.
In the Full Frame Mode observations, the source is highly off-center in the CCD image (Table \ref{tab:obs}), and bad pixels were found close to the source center region. 
For these observations, the event list was filtered using the selection expression \#XMMEA\_EP to exclude only photons which fall directly on the bad pixels. 
However, we did not exclude photon events located adjacent to the bad pixel (which normally would be excluded by the expression FLAG $=0$),
because when there is a bad pixel close to the center of the source region,
the effective area is evaluated more accurately with pixels around the bad
ones taken into account by the {\rm SAS} command {\tt arfgen} (\xmm help desk
2008, private communication).
In order to avoid events that affected multiple pixels, we used only single events (PATTERN$=0$) in the Full Frame Mode data.
Event lists thus extracted were input to {\tt ftool} {\tt grppha},
which grouped the events by at least 25 photons per bin. 
A systematic uncertainty of 2\%  was also appended to the output spectra using
{\tt grppha} in order to characterize the current level of calibration
accuracy.\footnote{See http://xmm.esac.esa.int/docs/documents/CAL-TN-0018.pdf,
(\it EPIC Status of Calibration and Data Analysis)}

The resultant spectra were fitted in {\rm XSPEC 12.3.0}\footnote{See
http://heasarc.gsfc.nasa.gov/docs/xanadu/xspec/} with the commonly used
photoelectrically absorbed blackbody plus power law model in the energy range
$0.6$-$12$ keV.
Because the hydrogen column density $N_H$ is not expected to be variable, we fixed this parameter for all the data sets and performed a joint fit.
The goodness of fit is reasonable (see $\chi^2_{\nu}$ in Table \ref{tab:par}).
The best-fit $N_H$ is $(1.012\pm0.007) \times 10^{22}\, \rm cm^{-2}$. 

This value is consistent with that estimated from fitting individual
absorption edges of elements O, Fe, Ne, Mg, and Si in the \xmm RGS spectra \citep{dv06b}.
The other parameters were set free to vary and their best-fit values are presented in Table \ref{tab:par}.
The best-fit blackbody temperature, blackbody radius, and power law index are plotted versus time in Figure \ref{fig:all}.

\begin{figure}
\includegraphics[scale=.9]{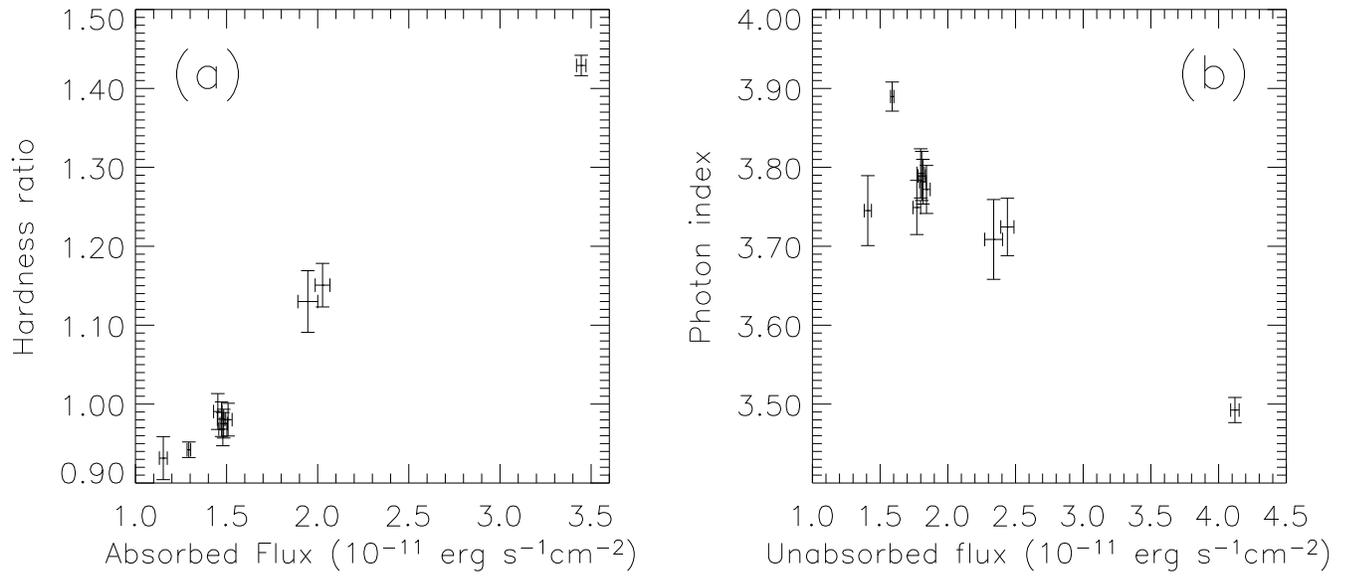}
\caption{\label{fig:flxhr}$(a)$ Hardness vs. absorbed flux. Hardness ratio
is defined as the ratio of 2--10 to 0.1--2 keV absorbed flux. $(b)$ Photon
index vs. 2--10 keV unabsorbed flux. }
\end{figure}

In order to look for correlations between spectral hardness and flux as
observed in other AXPs \citep{roz+05,cri+07,tgd+07,gdk+08}, we have looked for a correlation between hardness ratio and observed flux.
We define the hardness ratio to be the ratio of 2--10 keV absorbed flux to 0.1--2 keV absorbed flux.
We find the hardness ratio to be strongly correlated with the 2--10 keV
absorbed flux (as shown in Fig. \ref{fig:flxhr}{\em a}) in our observations.
An anti-correlation between photon index and 2--10 keV unabsorbed flux is also
seen, but has more scatter (as shown in Fig. \ref{fig:flxhr}{\em b}).
This is likely because the photon index is not a perfect measure of spectral hardness, as it can be strongly influenced by the spectral fit at the low end of the band.

\subsection{Pulsed fractions}
\label{sec:pf}
We folded the 0.1--2 and 2--10 keV light curves of each \xmm observation at the pulsar's period, determined using an ephemeris derived by phase coherent timing, from \xte monitoring (Table \ref{tab:obs}; see \citealt{dkg07} for details). 
Each pulse profile was constructed by folding the photons into 32 phase bins.
We measured area pulsed flux of the \xmm the same way we did for \xte (see
eq.\ref{eq:pf}), except that we used eight harmonics instead of six when smoothing
the light curves (for details see \citealt{adk08} in preparation).

The measured area pulsed fractions are plotted in Figure \ref{fig:all}{\em f}. 
A possible correlation between the 0.1--2 keV area pulsed fraction and the
2--10 keV unabsorbed flux is seen (Fig. \ref{fig:flxpf}, {\em filled circles}). 
A similar correlation was also found between the 0.1--2 keV area pulsed fraction and the 0.1--2 keV absorbed flux.
However, the correlation between 2--10 keV pulsed fraction and flux is not
significant (Fig. \ref{fig:flxpf}, {\em open boxes}).

We also measured the rms pulsed fraction from the profiles to compare with the area pulsed fraction results.  
The 2--10 keV rms pulsed fractions are consistent with being constant, while the 0.1--2 keV rms pulsed fractions have some variance, but no significant trend or correlation with other parameters.
The area and rms pulsed fractions are different by a factor that depends on the shape of the profile; as the pulse profile of \ttfn\, did change temporarily after the outburst (from a simple double peaked profile to triple peaked; \citealp{kgw+03,wkt+04}), the different result is not surprising.


\subsection{Flux evolution}
\begin{figure}
\includegraphics[scale=.8]{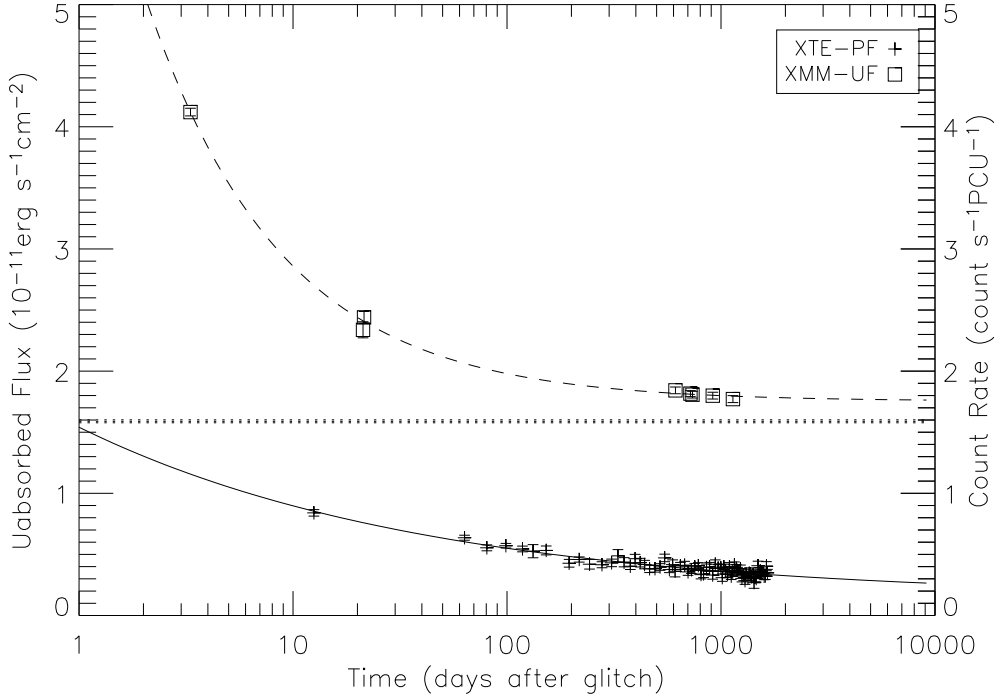}
\caption{\label{fig:flux} Evolution of \ttfn's 2--10 keV unabsorbed
phase-averaged flux ({\em squares})  measured from \xmm observations and 2--10 keV area
pulsed flux ({\em cross}) from \xte observations following its 2002 outburst.
\xmm unabsorbed fluxes are in units of $\rm 10^{-11} ergs~s^{-1}cm^{-2}$.
\xte area pulsed fluxes are in units of $\rm count~s^{-1}PCU^{-1}$.
The time axis is relative to the estimated glitch epoch (MJD 52,443.13).
The solid line is the power law plus constant model fit to the \xte area pulsed fluxes.
The dashed line is a fit of the same model, although having different best-fit
parameters, to the \xmm fluxes.
See Table \ref{tab:fit} for the best-fit parameters. 
The dotted line is the flux level in $\rm 10^{-11} ergs~s^{-1}cm^{-2}$ observed with \xmm one week before the outburst.
The uncertainty on this preoutburst flux is approximately the width of the line.
}
\end{figure}

We fit the unabsorbed fluxes measured in our \xmm observations after the
outburst with a power law plus constant decay model, $F(t)=F_b[(t-t_g)/({\rm
1~ day})]^{\alpha}+F_q$, where $F(t)$ denotes the unabsorbed flux, $F_b$ is
the unabsorbed source flux one day after the onset of the outburst, $F_q$ is
the quiescent flux and $t_g$ marks the glitch epoch MJD 52,443.13 \citep{wkt+04}.
A good fit of $\chi^2_{\nu}(\nu) =0.66(5)$ (Fig. \ref{fig:flux}, {\em dashed
line}) was found.
The best-fit power law index $\alpha=-0.69\pm0.03$ (Table \ref{tab:fit}).
The quiescent flux level we found from this power law fit is $(1.75 \pm
0.02)\times 10^{-11}\,\rm ergs\,s^{-1}cm^{-2}$, considerably higher than that
measured one week before the outburst [$(1.59\pm0.01)\times 10^{-11}\, \rm
ergs\,s^{-1}cm^{-2}$; Table \ref{tab:par}].
We also fit the \xmm unabsorbed fluxes with an exponential decay plus
quiescent level model, $F(t)=F_pe^{-(t-t_g)/\tau}+F_q$, where $F(t)$ is unabsorbed flux, $F_p$ is the peak flux, $F_q$ is the quiescent flux, $\tau$ is the decay timescale and $t_g$ marks the glitch epoch.
The fit is worse than that of the power law decay model but still acceptable, with $\chi^2_{\nu}(\nu)$ of $1.08(5)$. 
The best-fit decay timescale $\tau$ is $13.3\pm0.7$ days.
Best-fit flux decay parameters are presented in Table \ref{tab:fit}.

We also fit power law and exponential models to the area pulsed flux of
\ttfn\, measured by \xte from 12 to 1649 days after the glitch. 
A power law model fits the data
much better than the exponential model [$\chi^2_{\nu}(\nu)=1.18(69)$ for the
power law model, $\chi^2_{\nu}(\nu)=1.65(69)$ for the exponential decay model; Table \ref{tab:fit}],
which is evidence against the latter.
The best-fit exponential decay timescale for \xte data is $134\pm15$ days,
an order of magnitude different from the $\sim13$ day timescale found for the \xmm data. 

The best-fit power law plus constant model for the evolution of the \xte pulsed fluxes 
is different from that of the \xmm total fluxes.
This suggests that the 2--10~keV pulsed fractions were varying.
In principle, we can check this with the pulsed fraction measurements we made
with \xmm (see \S~\ref{sec:pf}).  Given the uncertainties on the \xmm 2--10~keV pulsed 
fractions (Table~\ref{tab:par}),
as well as those of the best-fit evolution models (Table~\ref{tab:fit}), 
we find that the two are in agreement.


\cite{gh07} fit the spectrum of \etten using a double-blackbody model when studying that source's relaxation following its outburst. 
In order to compare the spectrum and evolution of \ttfn\, to that of XTE~J1810$-$197, we also fit a photoelectrically absorbed double-blackbody model to \ttfn's spectra jointly.
A double-blackbody model does not fit the spectra as well as the blackbody plus power law model (see Table \ref{tab:par} for details). 
The best-fit $N_H$ for the double-blackbody model [$(0.568\pm0.03)\times
10^{22}\, \rm cm^{-2}$] is smaller than that from our blackbody plus power law
fit and is not consistent with the value  measured model independently from
RGS spectra [$(1.12\pm0.33)\times10^{22}\,\rm cm^{-2}$; \citealp{dv06b}],
but is consistent with the best-fit $N_H$ [$(0.5$-$0.7)\times10^{22}\,\rm cm^{-2} $] of CTB 109 measured by \cite{spg+04}. 

Unabsorbed fluxes obtained using the double-blackbody spectral model can also be fitted to a power law decay or an exponential decay model.
The best-fit power law index is $-0.73\pm0.04$, and the best-fit exponential timescale is $12.7\pm0.7$ days (Table \ref{tab:fit}), consistent with what we obtained using the blackbody plus power law spectral model. 
This indicates that our results for the decay parameters are independent of the choice of spectral model.
In the analysis of XTE~J1810$-$197 by \cite{gh07}, they found that both of the two-blackbody components' flux followed an exponential decay after \etten's 2003 outburst.
However, we find that the flux of \ttfn's soft blackbody component measured from our fourth and fifth observations (only $\sim$21 days after the outburst and glitch) were lower than that measured for the last five observations (see Table \ref{tab:par} for details).
This flux variation of the soft blackbody component therefore cannot be well fitted with an exponential or power law decay model.
The temperatures of both the hotter and cooler components were also lower in the fourth and fifth observations than in the last five observations.
The non-monotonic variation of the soft blackbody flux and the two components' temperature are different from what was observed by \cite{gh07} and suggest that the double-blackbody model is not a reasonable representation of the spectrum of \ttfn.
On the other hand, the spectral evolution from the blackbody plus power law spectral fit looks more reasonable. 
Using this spectral model, the blackbody radius in the first postoutburst
observation was small compared to that of the preoutburst observations and was
even smaller in the second and third postoutburst observations (Fig.
\ref{fig:all}{\em d}), suggesting that one or more hot spots formed after the outburst and were fading away in the next few months.
In the last five observations, the blackbody radius was as large as the preoutburst value, suggesting that the putative hot spots had completely faded away and the thermal radiation then mostly came from the bulk surface of the neutron star as it did before outburst.
Perhaps a more realistic spectral model such as that of \cite{gogk07,gog07} could describe the spectral evolution of \ttfn\, better, but such an analysis is outside the scope of this paper.
 
Based on \xte observations, \citet{wkt+04} found that the decay of \ttfn's
2002 outburst consisted of two parts:  a steeper power law decay in the
first few hours, and a slower power law decay afterwards.  They also
found that the total energy released (2--10 keV) in the slower decay was
$2.1\times10^{41}$ ergs, which is much larger than the total energy (2--60
keV) released in the bursts ($6\times10^{37}$ergs; \citealp{gkw04}).
We also studied the slower decay, by fitting a power law plus constant
model, instead of the simple power law model used by \citet{wkt+04}.
The total released energy, according to our best-fit model, is
roughly consistent with that calculated by \citet{wkt+04}:  we find
$\simeq3\times10^{41}$ ergs (2--10~keV), assuming that the outburst
will be over in 10000 days.  However, based on our best-fit exponential model,
the total energy released was somewhat smaller, $\simeq(3-4)\times10^{40}$ ergs (2--10 keV).

\section{Discussion}
\label{sec:disc}

In this paper, we have presented a comprehensive study of the X-ray recovery
of AXP 1E~2259+586 following its 2002 outburst.  Here we discuss the properties
of this recovery, compare them with those of other magnetar outbursts, and consider
how they constrain the magnetar model.

\subsection{Return to ``Quiescence''}

In our 2004 and 2005 {\it XMM} observations, the source's temperature
and unabsorbed fluxes were still higher than the preoutburst value
(Fig. \ref{fig:all}).  This suggests that the source was not fully back to
the preoutburst flux level.  Our power law fit to the flux decay shows
that the after-outburst quiescent flux level is $(1.75 \pm 0.02)\times
10^{-11}\, \rm ergs,s^{-1} cm^{-2}$, which is significantly higher than
the preoutburst value [$(1.59\pm0.01) \times 10^{-11}\,\rm ergs\,s^{-1}
cm^{-2} $; Table \ref{tab:fit}].  Either the 2005 flux had still not
returned to its quiescent level, or perhaps it had returned to quiescence
but the flux just before the event was unusually low.  Also possible is
that this (and other) AXPs do not have well-defined constant quiescent
fluxes, but have long-term flux variations.  Indeed, there is evidence for some
X-ray flux variability in \ttfn\, over the years since its discovery in 1981
\citep{bs96}.
Other AXPs also show variability on a variety of timescales 
(see \citealp{kas07} for a review).

\subsection{Comparison with other Magnetar Recoveries}

It is useful to compare the behavior observed from \ttfn\, with
that of other magnetars.  \ninet's flux was found to follow a
power law of index $-0.713\pm0.025 $ after its 1998 August 27 flare
\citep{wkg+01}.\footnote{Later the afterglow of the \ninet\, August 27 flare
was fitted with a power law plus constant model instead of the single
power law model used by \cite{wkg+01}, and a decay index of $\sim0.9$
was obtained \citep{fmw+03}.}  This has been interpreted as the cooling
of the magnetar outer crust following a sudden release of magnetic
energy \citep{let02}.  This model predicts a power law decay of index
$\sim -2/3$.  The flux of SGR 1627$-$41 was found to decay following a power law
of index $\sim-0.47$ since its 1998 source activation.  Approximately
800 days after the source activation, SGR 1627$-$41's flux suddenly
declined by a factor of 10.  This behavior is also well fitted by
the crust cooling model, although with some fine tuning \citep{kew+03}.
We fit the \xmm 2--10 keV unabsorbed fluxes of \ttfn\, with a power law
plus constant model, and found the best-fit power law index to be $-0.69
\pm 0.03 $, close to that of \ninet, and that predicted by the model.
This suggests that the \ttfn\, outburst afterglow may also be explained
by the diffusion of heat in the outer crust.

The transient AXP \etten exhibited an outburst in 2003. \cite{ims+04}
found that the afterglow of the \etten\, outburst as observed by \xte
could be described by a power law decay model ($F\propto t^{-\beta} $)
with $\beta=0.45-0.73 $.  This is similar to the behavior of \ttfn\,
and other SGRs.  However, \cite{gh07} found that, with more observations
taken by \chandra\, from 2003 to 2006, the afterglow of the \etten\,
outburst actually followed an exponential decay of timescale 233.5 days.
As we have shown in this paper, the pulsed and unabsorbed X-ray flux decay
of \ttfn\, favors the power law decay model over the exponential decay.
Perhaps the physical processes involved in the 2003 outburst of \etten
were different from those in 2002 outburst of \ttfn.

\subsection{Twisted Magnetosphere Model}

\cite{tlk02} reported that, if there exists a global twist of the
magnetosphere, the decay timescale $\tau$ of this twist would be 
\begin{equation} 
\label{eq:twist}
\tau = 40\Delta \phi^2 (\frac{L_X}{10^{35}{\rm
ergs\,s^{-1}}})^{-1}(\frac{B_{\rm pole}}{10^{14} \rm G})^2(\frac{R_{\rm
NS}}{10 \rm km})^3 \rm yr.  
\end{equation} 
\cite{wkt+04} argued that,
for \ttfn, the twist angle $\Delta \phi$ should be $\sim 10^{-2}$ rad.
Thus, the predicted twist relaxation timescale of \ttfn\, is several
hours, which is coincidently the timescale of the steeper flux decay
observed at the beginning of the afterglow.  

However, \citet{bt07} have 
shown more recently that this decay timescale is actually
expected to be much larger than equation~(\ref{eq:twist}) suggests.  This is because,
in their model, the self-induction of the twisted portion of the magnetosphere
accelerates particles from the stellar surface and initiates avalanches of pair
creation which forms the corona.  This corona persists in dynamic equilibrium,
maintaining the electric current, as long as dissipation permits.  The relevant
timescale in this picture for the decay of a sudden twist is given by
\begin{equation}
\tau \simeq 0.3 (\frac{L_X}{10^{35}{\rm ergs \, s^{-1}}}) (\frac{e\Phi_e}{\rm
GeV})^{-2}(\frac{R_{\rm NS}}{10 \rm km}) \,\, {\rm yr},
\end{equation}
where $L_X$ is the peak X-ray luminosity and $e\Phi_e$ is the voltage along the
twisted magnetic field lines and should nearly universally be $\sim$1~GeV \citep[see][]{bt07}.
For 1E~2259+586, we find $\tau
\simeq 1.2$~yr, given the peak luminosity $L_X \sim 4 \times 10^{35}(d/3
\; {\rm kpc})$~ergs~s$^{-1}$.  Thus, the longer observed decay after the
initial steep decline may indeed correspond to the untwisting of a coronal
flux tube in the \citet{bt07} picture, although the predicted timescale is somewhat 
smaller than the observed time to return to quiescence.  We note that the
\citet{bt07} model predicts a linear flux decline, in contrast
to what we have observed for 1E~2259+586 and what has been
observed for XTE~J1810$-$197 \citep{gh07}.  Moreover, in the $\sim$5 yr of {\it
RXTE} monitoring of 1E~2259+586 prior to its 2002 outburst \citep{gk02},
its pulsed X-ray luminosity in the 2--10~keV band was roughly constant at $\sim 2 \times
10^{34}$~ergs~s$^{-1}$.  This also is puzzling given the \citet{bt07}
prediction that if the time between large-scale events is longer than
the decay time from the previous event, the magnetar should enter a
quiescent state in which the observed luminosity is dominated by the
surface blackbody emission.  Why should the ``quiescent'' blackbody
emission from 1E~2259+586 be a full order of magnitude larger than that
from XTE~J1810$-$197, especially given the latter's much larger inferred
magnetic field ($1.7 \times 10^{14}$ versus $6 \times 10^{13}$~G)?
This disparity in ``quiescent,'' steady luminosities is even larger
when considering AXP 1E~1841$-$045, which has an apparently steady 2--10~keV
luminosity of $1.4 \times 10^{35}\rm ~ergs~s^{-1}$, and comparing with probable AXP
AX~1845$-$0258, which has quiescent luminosity approximately 2 orders of
magnitude smaller \citep{tkgg06}.  Distance uncertainties may contribute
but not on a scale that can significantly alleviate this problem.
This remains an interesting puzzle in magnetar physics.

The twisted magnetosphere or flux tube models generically predict that
the flux and spectral hardness of magnetars in outburst should be roughly
correlated due to increased scattering optical depth when the twist is
larger.  However, a similar prediction for a flux/hardness correlation was
made by \cite{og07} in their thermally emitting magnetar model, using a
simple prescription for the magnetosphere and scattering geometry, with
the latter stationary, i.e. invoking no variable magnetospheric twists.
\citet{gogk07} found that their model could reproduce the existing data
for XTE~J1810$-$197.  We note that hardness-intensity correlations have
now been observed for \seven\, \citep{cri+07}, \tfe\, \citep{tgd+07},
and as we report, in our \ttfn\, \xmm observations.  It would be interesting
to apply analysis of \cite{og07} to these data, but it is outside the scope of this paper.

\subsection{Other Observed Recovery Properties}

\begin{figure}
\includegraphics[scale=.8]{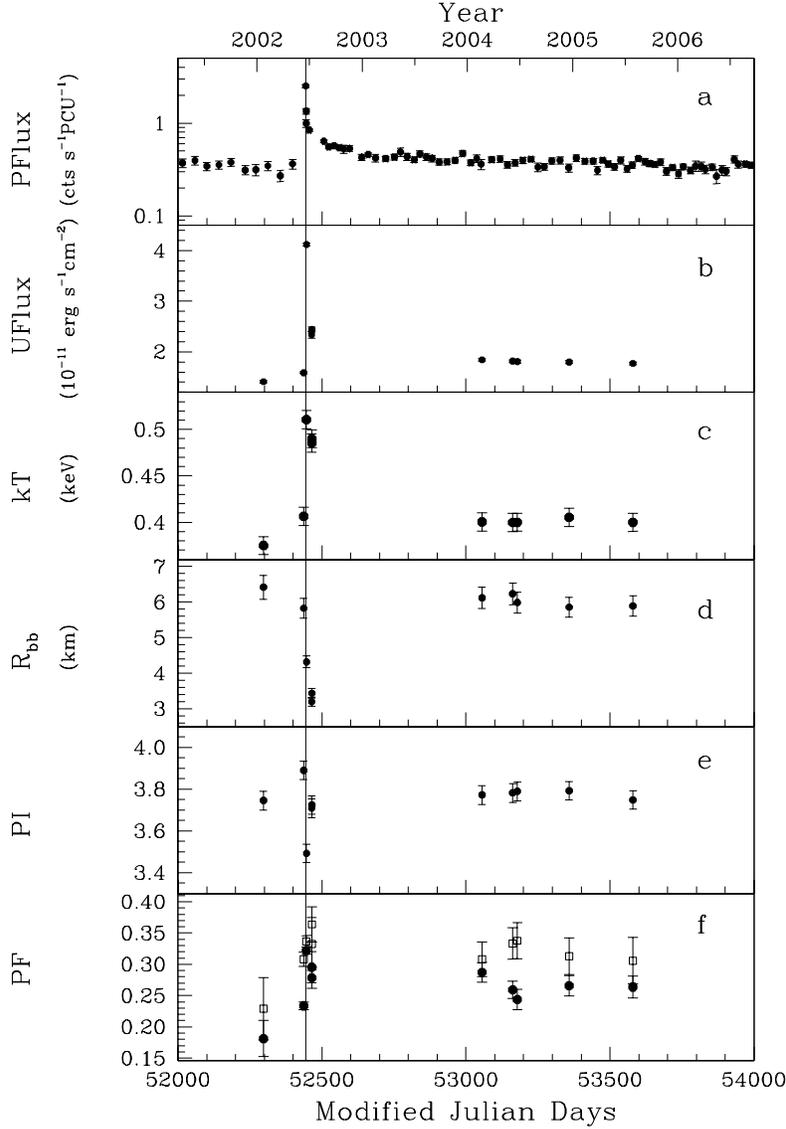}
\caption{\label{fig:all} Spectral and pulsed fraction evolution of \ttfn\,
during and following its 2002 outburst. ({\em a}) 2--10 keV
area pulsed flux measured in \xte monitoring observations; ({\em b}) 2--10 keV
unabsorbed phase-averaged flux from \xmm observations (all lower panels are
also from \xmm observations);  ({\em c}) blackbody temperature ($kT$), ({\em
d}) blackbody radius; ({\em e}) photon index; ({\em f}) 0.1--2 keV area
pulsed fraction ({\em filled circles}) and 2--10 keV area pulsed fraction
({\em open boxes}). 
A distance of 3 kpc \citep{kuy02} is assumed to calculate the blackbody radius.
The vertical line denotes the 2002 glitch epoch, MJD 52,443.13 \citep{wkt+04}.
}
\end{figure}

The fact that the rms and area pulsed fractions remained largely constant while the
blackbody radius (in the blackbody plus power law model) changed by
a factor of $\sim$2 (Fig. \ref{fig:all}) is worth considering, if the
empirical blackbody plus power law spectrum model somehow resembles the
real radiation mechanism.  
Pulsed fraction should generally decrease when the thermally radiating
region on the star grows, provided that this region is not very small
compared to the entire surface.  Any realistic spectral model which
takes radiative transfer in the atmosphere and scattering through the
magnetosphere into account should be able to reproduce the observation
in this regard as well.

\begin{figure}
\includegraphics[scale=0.6]{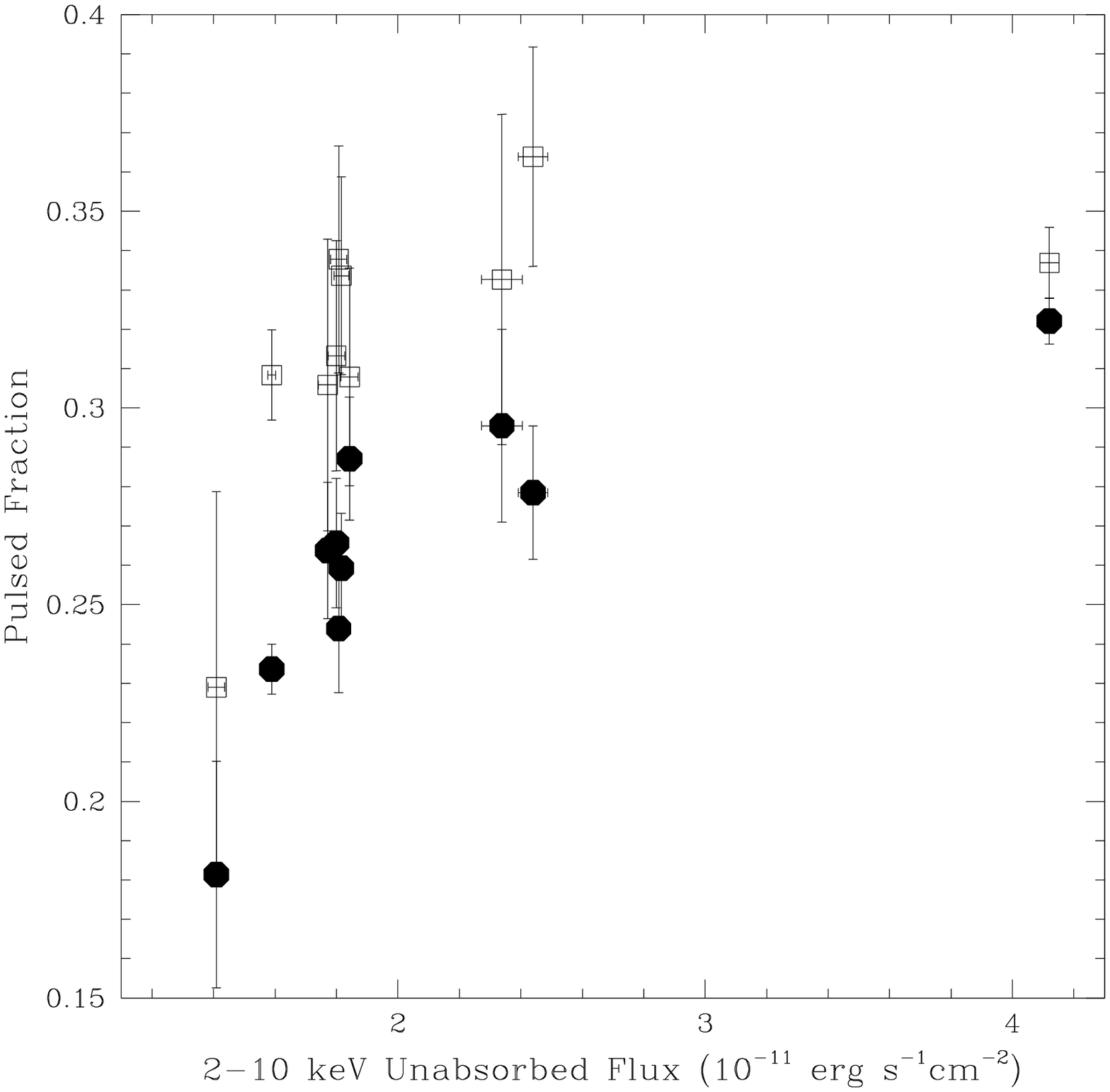} \\ 
\caption {\label{fig:flxpf} The 2--10 keV area pulsed fraction ({\em open
boxes}) vs. 2--10 keV unabsorbed flux ; 0.1--2 keV area pulsed fraction ({\em
filled circles}) vs. 2--10 keV unabsorbed flux.} 
\end{figure}

A clear anti-correlation between  \tfe's pulsed fraction and unabsorbed
flux has been observed \citep{tmt+05,gkw06,tgd+07}.  
However, we found no such correlation in the 2--10 keV band for \ttfn.  
On the contrary, its 0.1--2 keV area pulsed fractions seem to be correlated with
both 0.1--2 and 2--10 keV unabsorbed fluxes (see Fig. \ref{fig:flxpf}).
 \cite{gh07} found that \etten's pulsed
fraction measured between 2003 and 2006 after its outburst decreased
with the decay of its flux, i.e. \etten's pulsed fraction is also
correlated with flux.  Thus, the striking anti-correlation between pulsed
fraction and flux observed from \tfe\, is clearly not universal.

Finally, we note that the near-infrared flux decay of \ttfn\, was found
to follow a power law of index $-0.75^{+0.22}_{-0.33}$ when fitted to
a power law plus constant model \citep{tkvd04}.  This decay index is
close to what we found for the X-ray flux decay, thus confirming the
reported correlation between near-IR and X-ray fluxes postoutburst.\footnote{The $-$0.22 X-ray decay index reported by \cite{wkt+04} and
the $-$0.21 near-infrared flux decay index reported by \cite{tkvd04} were
obtained from a simple power law fitting, i.e. with no quiescent level
included in the fit.}  \citet{tgd+08} and \citet{wbk+08} showed that
the near-IR flux of 1E~1048.1$-$5937 do show correlation with X-rays at
times of outbursts.  However, \citet{crp+07} show that the near-IR flux
variation of XTE~J1810$-$197 is not simply correlated with X-ray flux
nor even monotonic postoutburst.  Thus, the AXP picture with regard to
near-IR variability is not yet fully clear.

\acknowledgements
We thank A. Beloborodov, A. Cumming, and G. Vasisht for useful discussions.
Support for this work was provided by an NSERC Discovery grant Rgpin
228738-03, an R. Howard Webster Fellowship of the Canadian Institute for
Advanced Research, Les Fonds de la Recherche sur la Nature et les
Technologies, a Canada Research Chair, and the Lorne Trottier Chair in Astrophysics and Cosmology to VMK.  
RD was supported by an NSERC PGSD scholarship.

\bibliographystyle{apj}
\bibliography{myrefs,journals1,modrefs,psrrefs,crossrefs}

\clearpage

\begin{deluxetable}{ccccccc}
\tabletypesize{\footnotesize}
\tablecaption{\label{tab:obs} \xmm observation log for 1E~2259$+$586.} 
\tablehead{
 Name & \xmm  &  Date & Date & pn on-times \tablenotemark{a} & Off-axis angle  & Frequencies \tablenotemark{b}  \\
      & Obsid             &  (MJD TDB)  & (YY-MM-DD) & (ksec)   & (arcmin)  & (s$^{-1}$)               
}
\startdata
 Obs1\tablenotemark{c \rm d}	& 0057540101  &   52,296.791 & 02-01-22 & 8.5   &   8.7 & 0.1432871204(7)   \\
 Obs2\tablenotemark{c}		& 0038140101  &   52,436.413 & 02-06-11 & 26.3  &   2.0 & 0.143287000(9)   \\
 Obs3				& 0155350301  &   52,446.446 & 02-06-21 & 16.4  &   2.0 & 0.14328759(1)  \\
 Obs4\tablenotemark{d}		& 0057540201  &   52,464.368 & 02-07-09 & 5.2   &  10.7 & 0.14328754(1)   \\
 Obs5\tablenotemark{d}		& 0057540301  &   52,464.602 & 02-07-09 & 10.2  &  10.3 & 0.14328754(1)    \\
 Obs6				& 0203550301  &   53,055.596 & 04-02-20 & 3.6   &   1.9 & 0.143286974(7)  \\
 Obs7				& 0203550601  &   53,162.655 & 04-06-06 & 4.8   &   2.0 & 0.143286882(4)  \\
 Obs8				& 0203550401  &   53,178.634 & 04-06-22 & 3.4   &   2.0 & 0.143286868(2)  \\
 Obs9				& 0203550501  &   53,358.014 & 04-12-19 & 3.5   &   2.0 & 0.143286714(1)  \\
 Obs10				& 0203550701  &   53,579.970 & 05-07-28 & 3.3   &   1.9 & 0.143286523(7)  \\

\enddata
\tablenotetext{a}{On-times quoted reflect on-source times after filtering of background flares.}
\tablenotetext{b}{Frequencies are from contemporaneous \xte observations. }
\tablenotetext{c}{Observations taken before the outburst, MJD 52,443.13 \citep{wkt+04}. }
\tablenotetext{d}{These three observations were taken in extended Full Frame Mode; all the others were taken in Small Window Mode. }
\end{deluxetable}

\begin{deluxetable}{lllllllllll}
\tablewidth{495pt}
\tabletypesize{\scriptsize}
\tablecaption{\label{tab:par} 1E~2259$+$586's best-fit spectral parameters and pulsed fractions. } 
\tablehead{
 Parameter\tablenotemark{ a}  \tablenotemark{b}    &   Obs1     &  Obs2     &   Obs3    &   Obs4     &  Obs5   &  Obs6     &  Obs7     & Obs8    &   Obs9     &  Obs10  
}
\startdata
\multicolumn{11}{c}{Blackbody plus power law model}  \\ \hline
 
$N_{H}$ (10$^{22}$ cm$^{-2}$)	& 1.012(7) & 1.012(7) & 1.012(7) & 1.012(7) & 1.012(7) & 1.012(7) & 1.012(7) & 1.012(7) & 1.012(7)& 1.012(7)\\
$kT$ (keV)			& 0.37(1) & 0.406(2)  & 0.510(4) & 0.48(2) & 0.49(1) & 0.400(7) & 0.400(5) & 0.400(7) & 0.405(7) & 0.400(7)\\
$\Gamma$			& 3.75(4)  & 3.89(2)  & 3.49(2)  & 3.71(5)   & 3.72(4) & 3.77(3) & 3.78(3) & 3.79(3) & 3.79(3) & 3.75(3)  \\
Flux\tablenotemark{c} & 1.15(2) & 1.29(1) & 3.45(3) & 1.95(5) & 2.03(4) & 1.51(2) & 1.48(2) & 1.48(2) & 1.47(2) & 1.45(2) \\
Unabs Flux\tablenotemark{d} & 1.41(3) & 1.59(1) & 4.12(3) & 2.34(7) & 2.44(5) & 1.84(3) & 1.82(2) & 1.81(3) & 1.80(3) & 1.77(3) \\
PL/BB ratio\tablenotemark{e} & 1.8(3) & 1.2(2) & 1.6(2) & 2.6(4) & 2.1(3) & 1.6(2) & 1.5(2) & 1.7(2) & 1.5(2) & 1.7(2) \\
Hardness \tablenotemark{f} & 0.93(3) & 0.94(1) & 1.43(1) & 1.13(4) & 1.15(4) & 0.98(2) & 0.98(2) & 0.97(2) & 0.98(2) & 0.99(2) \\
$\chi^2_{\nu}(\nu)$                & \multicolumn{2}{l}{ 1.02(5800)}   &\multicolumn{2}{c}{ ($P =0.12$)\tablenotemark{g} }  &  &  &  &  &  &  \\
\hline
\multicolumn{11}{c}{Double blackbody model}  \\ \hline
$N_{H}$ (10$^{22}$ cm$^{-2}$)	& 0.568(3) & 0.568(3) & 0.568(3) & 0.568(3) & 0.568(3) & 0.568(3) & 0.568(3) & 0.568(3) & 0.568(3)& 0.568(3)\\
Cooler $kT$ (keV)		& 0.362(5) & 0.372(2)  & 0.390(3) & 0.330(7) & 0.335(5) & 0.371(3) & 0.380(3) & 0.371(3) & 0.370(4) & 0.371(3)\\
Hotter $kT$ (keV)		& 0.77(4)  & 0.82(1)  & 0.86(1)  & 0.74(3)   & 0.73(2) & 0.85(3) & 0.94(2) & 0.89(3) & 0.85(3) & 0.86(3)  \\
Flux\tablenotemark{c} & 1.12(6) & 1.27(2) & 3.34(4) & 1.85(9) & 1.93(7) & 1.48(4) & 1.46(3) & 1.44(4) & 1.44(4) & 1.41(4) \\
Unabs Flux\tablenotemark{d} & 1.26(6) & 1.42(2) & 3.69(4) & 2.05(11) & 2.15(8) & 1.65(5) & 1.63(4) & 1.62(5) & 1.61(5) & 1.58(5) \\
HB/CB ratio\tablenotemark{h} & 0.8(1) & 0.6(1) & 1.2(2) & 2.1(3) & 2.0(3) & 0.7(1) & 0.53(7) & 0.64(9) & 0.7(1) & 0.7(1) \\
Hardness \tablenotemark{f}  & 0.91(5) & 0.92(1) & 1.37(2) & 1.07(6) & 1.09(4) & 0.96(3) & 0.95(2) & 0.94(3) & 0.95(3) & 0.96(3) \\
$\chi^2_{\nu}(\nu)$                & \multicolumn{2}{l}{ 1.11(5800)}   &\multicolumn{2}{c}{($P=4.6\times10^{-9}$)\tablenotemark{g} }    &  &  &  &  &  &  \\
\hline
\multicolumn{11}{c}{Pulsed fractions}  \\ \hline
PF(0.1--2 keV)\tablenotemark{i} & 0.18(3) & 0.234(6) & 0.322(6) & 0.30(2) & 0.28(2) & 0.29(2) & 0.26(1) & 0.24(2) & 0.27(2) & 0.26(2)\\ 
PF(2--10 keV)\tablenotemark{i} & 0.23(5) & 0.30(1) & 0.339(9) & 0.33(4) & 0.36(3) & 0.30(3) & 0.33(2) & 0.34(3) & 0.31(2) & 0.29(3)\\ 

\enddata
\tablenotetext{\rm a}{ Numbers in parentheses indicate the 1$\sigma$ uncertainty in the least significant digit. Note that these
uncertainties reflect the 1$\sigma$ error for a reduced $\chi^2$ of unity. }
\tablenotetext{\rm b}{Best-fit parameters from a joint fit to all data sets. $N_H$ in all data sets was set to be the same; other parameters were allowed to vary from observation to observation.}
\tablenotetext{\rm c} {(10$^{-11}$ ergs s$^{-1}$cm$^{-2}$). Observed flux from
both spectral components in the range 2--10 keV. }
\tablenotetext{\rm d} {(10$^{-11}$ ergs s$^{-1}$cm$^{-2}$). Unabsorbed flux
from both spectral components in the range 2--10 keV. }
\tablenotetext{\rm e} {The ratio of power law flux to blackbody flux in the 2--10 keV band (corrected for absorption).}
\tablenotetext{\rm f} {Spectral hardness defined as the ratio of 2--10 keV
absorbed flux to 0.1--2 keV absorbed flux.}
\tablenotetext{\rm g} {The probability for the $\chi^2_{\nu}$ to be higher
than that was observed, assuming the model is correct.}
\tablenotetext{\rm h} {The ratio of hot blackbody flux to cool blackbody flux in the 2--10 keV band (corrected for absorption).}
\tablenotetext{\rm i} {The area pulsed fractions. }
\end{deluxetable}


\begin{deluxetable}{lcccll}
\tabletypesize{\scriptsize}
\tablecaption{Best-fit parameters for the \ttfn\, flux decay
  \label{tab:fit} }
\tablehead{
\colhead{Power law decay\tablenotemark{a}} & \colhead{$F_q (10^{-11}\rm
ergs\,s^{-1}cm^{-2})$} & \colhead{$F_b(10^{-11}\rm ergs\,s^{-1}cm^{-2})$}  & \colhead{$\alpha$} &
\colhead{$\chi^2$} & \colhead{$\chi^2/\nu$} 
}
\startdata
\xmm UF (BB+PL) \tablenotemark{b}&  $(1.75 \pm 0.02)$   & $(5.40 \pm 0.21)$ &  $-0.69 \pm 0.03$ & 3.31 & 0.66\\
\xmm UF (BB+BB) \tablenotemark{c}&  $(1.58 \pm 0.01)$   & $(5.07 \pm 0.22)$ &  $-0.73 \pm 0.04$ & 1.22 & 0.24\\
\xte PF \tablenotemark{d}&$(0.14 \pm 0.06)\rm cts~s^{-1} PCU^{-1}$ & $(1.4 \pm 0.1)\rm cts~s^{-1} PCU^{-1}$ &  $-0.27 \pm 0.05$ & 81.3 & 1.18\\
\hline
\colhead{Exponential decay\tablenotemark{e}} & \colhead{$F_q(10^{-11}\rm
ergs\,s^{-1}cm^{-2})$ } & \colhead{$F_p(10^{-11}\rm ergs\,s^{-1}cm^{-2})$
 }  & \colhead{$\tau$ (days) }&
\colhead{$\chi^2$} & \colhead{$\chi^2/\nu$} \\ 
\hline
\xmm UF (BB+PL) \tablenotemark{b}&  $(1.81 \pm 0.01)$   & $(2.97 \pm 0.06)$ &  $13.3 \pm 0.7$ & 5.39 & 1.08\\
\xmm UF (BB+BB) \tablenotemark{c}&  $(1.62 \pm 0.01)$   & $(2.69 \pm 0.05)$ &  $12.7 \pm 0.7$ & 1.737 & 0.35\\
\xte PF \tablenotemark{d}&$(0.362 \pm 0.005 )\rm cts~s^{-1} PCU^{-1}$ & $(0.46 \pm 0.03)\rm cts~s^{-1} PCU^{-1}$ &  $134 \pm 15$ & 113.5 & 1.65\\
\enddata
\tablenotetext{\rm a}{Power law decay model defined as $F(t)=F_b((t-t_g)/({\rm
1~day}))^{\alpha}+F_q$, where $F(t)$ is unabsorbed flux, $F_q$ is the
quiescent flux, $\alpha$ is the power law index and $t_g$ is the glitch epoch
MJD 52,443.13.}
\tablenotetext{\rm b}{\xmm unabsorbed flux decay measured using a blackbody plus power law spectral model.}
\tablenotetext{\rm c}{\xmm unabsorbed flux decay measured using a double-blackbody spectral model.}
\tablenotetext{\rm d}{\xte area pulsed flux in units of {$\rm cts~s^{-1} PCU^{-1}$}.}
\tablenotetext{\rm e}{Exponential decay model, defined as
$F(t)=F_pe^{-(t-t_g)/\tau}+F_q$, where $F(t)$ is unabsorbed flux, $F_p$ is the
peak flux, $F_q$ is the quiescent flux, $\tau$ is the decay timescale and
$t_g$ is the glitch epoch MJD 52,443.13.}
\end{deluxetable}

\clearpage

\clearpage

\clearpage

\clearpage

\end{document}